\documentclass[conference,a4paper]{IEEEtran}
\usepackage[pdftex]{graphicx}
\ifCLASSINFOpdf
\else
\fi
\usepackage{amsmath}
\ifCLASSOPTIONcompsoc
  \usepackage[caption=false,font=normalsize,labelfont=sf,textfont=sf]{subfig}
\else
  \usepackage[caption=false,font=footnotesize]{subfig}
\fi
\usepackage{url}

    \DeclareMathOperator*{\argmin}{arg\,min}
    \DeclareMathOperator*{\argmax}{arg\,max}
    \DeclareMathOperator*{\sargmax}{arg\,max^*}

\usepackage{comment}
\usepackage{balance}

\IEEEoverridecommandlockouts

\begin{document}
%
\title{Analysis of Home Location Estimation\\with Iteration on Twitter Following Relationship}


\author{
    \IEEEauthorblockN{Shiori Hironaka \quad Mitsuo Yoshida \quad Kyoji Umemura}
    \IEEEauthorblockA{Department of Computer Science and Engeneering,\\
    Toyohashi University of Technology,\\
    Toyohashi, Aichi, Japan\\
    s143369@edu.tut.ac.jp, yoshida@cs.tut.ac.jp, umemura@tut.jp}
}


%


\IEEEpubid{\makebox[\columnwidth]{978--1--5090--1636--5/16/\$31.00~\copyright~2016
IEEE \hfill}\hspace{\columnsep}\makebox[\columnwidth]{\hfill }}

\maketitle

\begin{abstract}
User's home locations are used by numerous social media applications, such as social media analysis.
However, since the user's home location is not generally open to the public, many researchers have been attempting to develop a more accurate home location estimation.
A social network that expresses relationships between users is used to estimate the users' home locations.
The network-based home location estimation method with iteration, which propagates the estimated locations, is used to estimate more users' home locations.
In this study, we analyze the function of network-based home location estimation with iteration while using the social network based on following relationships on Twitter.
The results indicate that the function that selects the most frequent location among the friends' location has the best accuracy.
Our analysis also shows that the 88\% of users, who are in the social network based on following relationships, has at least one correct home location within one-hop (friends and friends of friends).
According to this characteristic of the social network, we indicate that twice is sufficient for iteration.
\end{abstract}

\renewcommand\IEEEkeywordsname{Keywords}
\begin{IEEEkeywords}
network-based home location estimation; label propagation; Twitter
\end{IEEEkeywords}

%
\IEEEpeerreviewmaketitle

\section{Introduction}

Social media provides content with user data, and user data is used to analyze the content of social media.
Based on this feature,
the user's attributes are used for social media analysis and numerous applications.
There are various user attributes such as gender, hobbies, and the user's home location.
Among the various user attributes, this study focuses on the user's home location.
However, since the user's home location is not generally open to the public,
many researchers have been attempting to develop a more accurate home location estimation.

There are three ways of location estimation mainly: the content-based method (e.g., \cite{Cheng2010,Mahmud2014}), the network-based method~\cite{Jurgens2015}, and the hybrid method (e.g., \cite{Li2012}) that is based on both of content and network.
The network-based method has an advantage that does not depend on languages.
In this study, we focus on the network-based home location estimation.

The network-based home location estimation is a method that can estimate a user's home location by using a social network that expresses relationships between users.
The first approach for network-based location estimation is using the home locations of friends~\cite{Backstrom2010,DavisJr2011,McGee2013}.
The next approach is to estimate more home locations by using home location estimation with iteration~\cite{Jurgens2013,Kong2014}, in order to improve the estimation coverage using the first approach.
The social network is generally created from the \emph{following relationships} on Twitter\footnote{\url{https://twitter.com/}} (follow-follower relationship) between users.
However,
we need to know the performance of network-based home location estimation with iteration using the social network based on following relationship.

In this study, we analyze the function of network-based home location estimation with iteration while using the social network based on following relationships on Twitter.
According to the characteristic of a social network on Twitter,
twice is described as a sufficient iteration count in the social network based on the following relationships.

\section{Spatial Label Propagation}

\IEEEpubidadjcol

In the network-based location estimation method, the home location of users is estimated based on the social network and a few parts of the assigned user's home locations.
We define the social network as a simple directed graph, the node as a user, and the edge as the relationship between the users.
The user's home location is expressed as a label assigned to the node.

Spatial label propagation (SLP)~\cite{Jurgens2013} imports the idea of label propagation~\cite{Zhu2003} to network-based home location estimation.
SLP is a way of applying the home location estimation method using the labels of the adjacent nodes iteratively.
SLP is capable of estimating a greater number of nodes, and performs well in a sparse graph.
In addition, SLP can estimate home locations using labels of non-adjacent nodes.

SLP has two parameters, the \emph{select function} and the \emph{iteration count}.
We use the network-based home location estimation method using adjacent labels as the select function.
In this study, we use four network-based home location estimation methods using adjacent labels as the select function, explained in Section~\ref{sec:pbm} to \ref{sec:rn}.
The select function of node $u$ defined as $Select(u)$ uses only the following information:
$N_u$ is a set of adjacent nodes of node $u$,
and a set of labels of $N_u$.

We use the following variables for explanation:
$L$ is a set of learning data (nodes),
$N_u$ is a set of adjacent nodes of node $u$,
$A$ is a set of labels (areas),
$l_u$ is a correct label of node $u$,
$dist(a,b)$ is the distance between labels $a$ and $b$.
The distance between labels is the geographic distance between two centroids of areas corresponding with the labels calculated by Hubeny's distance formula.

\subsection{Probability Model} \label{sec:pbm}

Probability Model~\cite{Backstrom2010} is the method that selects the label (home location) having the highest likelihood of the target node, with the model of the probability that there is an edge at the geographic distance.
%
%
When the geographic distance between the nodes is $d$, the probability of the edge being present $p(d)$ is expressed in \eqref{eq:pmodel},
where $a$, $b$ and $c$ are the static parameters of a real number.
We use $a=0.0019$, $b=0.196$, $c=-1.05$ in original paper~\cite{Backstrom2010}.
Home location of the node $u$ is estimated by \eqref{eq:poptimized} using the probability model\footnote{We make up for $\gamma_l(l)$ to equation $\gamma(l, u)$ because it is thought that the equation used in the original paper is incorrect.}.
\begin{gather}
p(d) = a(d + b)^c \label{eq:pmodel} \\
\gamma_l(l) = \prod_{v \in L} [ 1 - p(dist(l, l_v)) ] \nonumber \\
\gamma(l, u) = \prod_{v \in N_u \cap L} \frac{p(dist(l, l_v))}{1 - p(dist(l, l_v))} \gamma_l(l) \nonumber \\
ProbabilityModel(u) = \argmax_{l \in \{ l_n | n \in N_u \cap L \} } \gamma(l, u) \label{eq:poptimized}
\end{gather}

\subsection{Majority Vote}

Majority Vote~\cite{DavisJr2011} is the method that selects the most frequent label (home location) among the labels of the adjacent nodes (friends).
The hypothesis based on this method is that the majority of my friends live the same location with me.
In the original paper~\cite{DavisJr2011}, the concept of majority may be ambiguous.
The paper does not specify the majority when two or more labels are the most frequent in the labels of adjacent nodes.
Therefore, in this study, we preferentially choose a label which appears frequently in the entire social network.
The method is expressed in \eqref{eq:majorityvote},
where $\sargmax$ is defined that returns a set of the equivalent.
\begin{gather}
S_u = \sargmax_{ l \in \{ l_n | n \in N_u \cap L \} } |\{ v|v \in N_u \cap L, l = l_v \}| \nonumber \\
MajorityVote(u) = \argmax_{l \in S_u} |\{ n | n \in L, l = l_n \}| \label{eq:majorityvote}
\end{gather}

This method contains two parameters, the range of the number of adjacent nodes and the minimum number of votes.
Since other methods do not have the corresponding parameters, the maximum range is selected for the Majority Vote method.
The range of the number of adjacent nodes of the user is zero to infinity, and the minimum number of votes is zero.

\subsection{Geometric Median}

Geometric Median~\cite{Jurgens2013} is the method that selects the geometric median among the labels of the adjacent nodes.
The method is expressed in \eqref{eq:geometricmedian}.
\begin{equation}
\begin{split}
&GeometricMedian(u) \\
=& \argmin_{l \in \{ l_n | n \in N_u \cap L \} } \sum_{ x \in N_u \cap L, n \ne x } dist(l, l_x) \label{eq:geometricmedian}
\end{split}
\end{equation}

\subsection{Random Neighbor} \label{sec:rn}

Random Neighbor is the baseline method that selects randomly among the labels of the adjacent nodes.
This method is expressed in \eqref{eq:randomneighbor},
where $choice(S)$ is the function that selects an element randomly from the set $S$.
\begin{equation}
RandomNeighbor(u) = l_{choice(N_u \cap L)} \label{eq:randomneighbor}
\end{equation}

\section{Dataset}

In this study, we analyze home location estimation of Twitter users using the social network based on following relationships.
We use home location data and their social network for the experiment, and we describe the details below.
Specifically, the home location data contains 52,508 users, and the social network includes 8,003,858 nodes and 40,453,444 edges.

\subsection{Home Location from Geo-tagged Tweets}

In this study, a home location is an area of the city, as expressed in the previous study~\cite{DavisJr2011}.
We find a city where the coordinate of the geo-tagged tweet is included,
and we use the city as the area.
%
Furthermore,
we assume that a user is active mainly around their home location.
We assign the home location as the most frequent city where the user posts geo-tagged tweets at least five times to the user who posts geo-tagged tweets more than 365 times.

We collected 250,564,317 geo-tagged tweets posted in the rectangle that includes Japan\footnote{Range of lat. 20$^\circ$ to 50$^\circ$N and long. 110$^\circ$ to 160$^\circ$E.} in 2014,
using the Twitter Streaming API\footnote{\url{https://dev.twitter.com/streaming/reference/post/statuses/filter}}.
We excluded tweets posted by BOT accounts.
As a result, we assigned home locations to 71,166 users.

\subsection{Social Network based on Following Relationship}\label{sec:making}

Twitter users can follow other users by subscribing to the users' tweets.
A followee is the user who is following, and a follower is the user who follows the user.
The followee and follower relationships are called \emph{following relationships}.
In this study, we create a social network based on following relationships.
At first,
we collect two sets of following relationships of the users to which a home location can be assigned; followees of the user\footnote{\url{https://dev.twitter.com/rest/reference/get/friends/ids}} and followers of the user\footnote{\url{https://dev.twitter.com/rest/reference/get/followers/ids}}.
Then, we merge the two sets,
and create a social network by making an edge when the users follow one another mutually.

We collected the following relationships among the users who were assigned a home location.
We were able to collect the following relationships of 52,508 users among the 71,166 users in July 2015,
and their social network had 8,003,858 nodes and 40,453,444 edges.

\section{Experiment and Discussion}

In this section, we conduct experiments to compare the select functions.
We first describe the experimental setup.
Then, we present results and analysis, and provide some discussion regarding those results.

\subsection{Experimental Setup}

The evaluation is based on a 10-fold cross-validation with 52,508 users who were assigned a home location.
We compare four select functions with the social network created by the following relationships.

The experimental results are reported with their precision, recall and F1 (f-measure).
We have defined precision as the ratio of users that have been correctly estimated a home location,
recall as the ratio of users that have been correctly estimated a location to the users in the test dataset,
and F1 as the harmonic mean of precision and recall.
In addition, we use the coverage that shows the ratio of users who can be estimated for analysis.
Because isolated nodes exist in the social network, the maximum coverage is not 100\%.
%
We also evaluate by mean error distance and median error distance based on the distance between correct and estimated locations.


The calculation of six metrics can be seen in \eqref{eq:metric},
where $X$ is the set of estimated users,
$T$ is the set of test users,
$l_u$ is the home location of user $u$,
$e_u$ is the estimated location of user $u$,
$D_U$ is the set of distance $dist(l_u, e_u)$ of all users $u \in U$,
$mean(A)$ is the function that returns the mean value in the set $A$,
and $median(A)$ is the function that returns the median value in the set $A$.
The evaluation value is the mean of the metrics for each fold.
\begin{equation}
\begin{split}
\label{eq:metric}
Precision(T, X) = \frac{ |\{ u | u \in T \cap X, l_u = e_u \}| }{ | T \cap X | } \\
Recall(T, X) = \frac{ |\{ u | u \in T \cap X, l_u = e_u \}| }{ |T| } \\
F1(T, X) = \frac{ 2*Precision(T,X)*Recall(T,X) }{ Precision(T,X)+Recall(T,X) } \\
Coverage(T, X) = \frac{ |T \cap X| }{ |T| } \\
    MeanErrorDistance(T, X) = mean(D_{T \cap X}) \\
    MedianErrorDistance(T, X) = median(D_{T \cap X})
\end{split}
\end{equation}

\subsection{Results and Analysis}

\begin{figure}[!tp]
\centering
\subfloat[Precision]{%
	\includegraphics[width=0.99\linewidth]{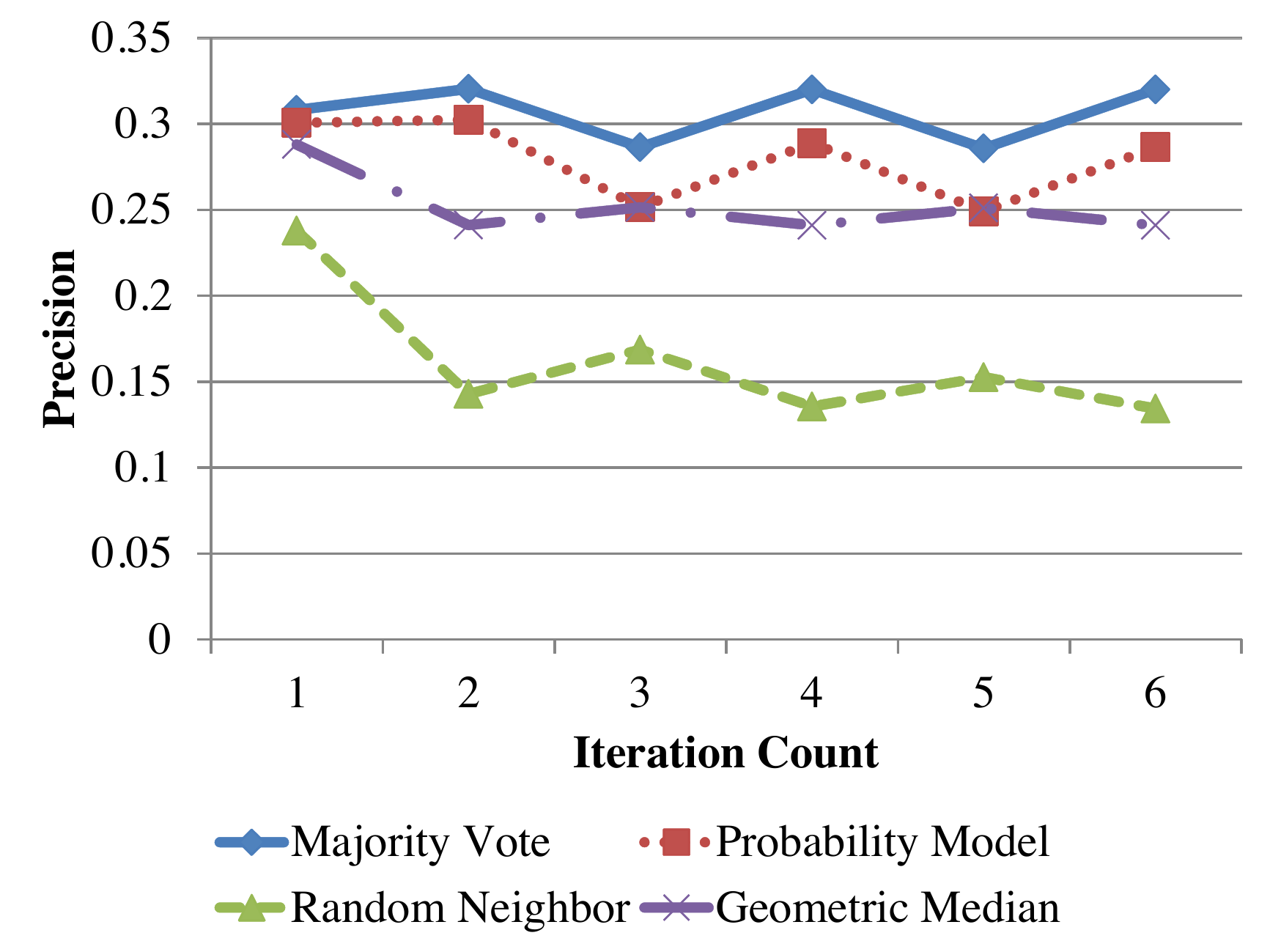}%
	\label{fig:slp-mutual-precision}%
}
\hfil
\subfloat[Recall]{%
	\includegraphics[width=0.99\linewidth]{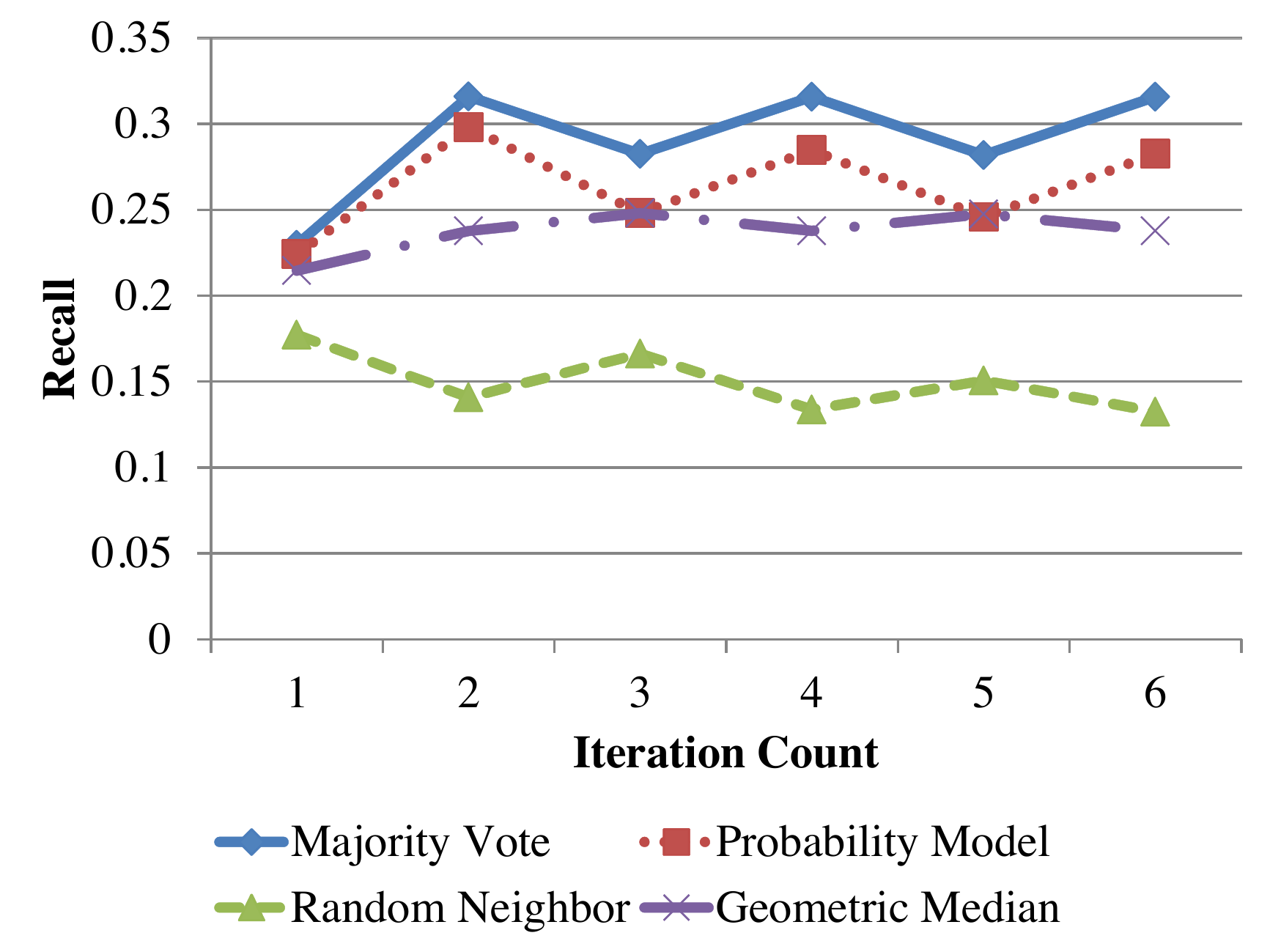}%
	\label{fig:slp-mutual-recall}%
}
\hfil
\subfloat[F1]{%
	\includegraphics[width=0.99\linewidth]{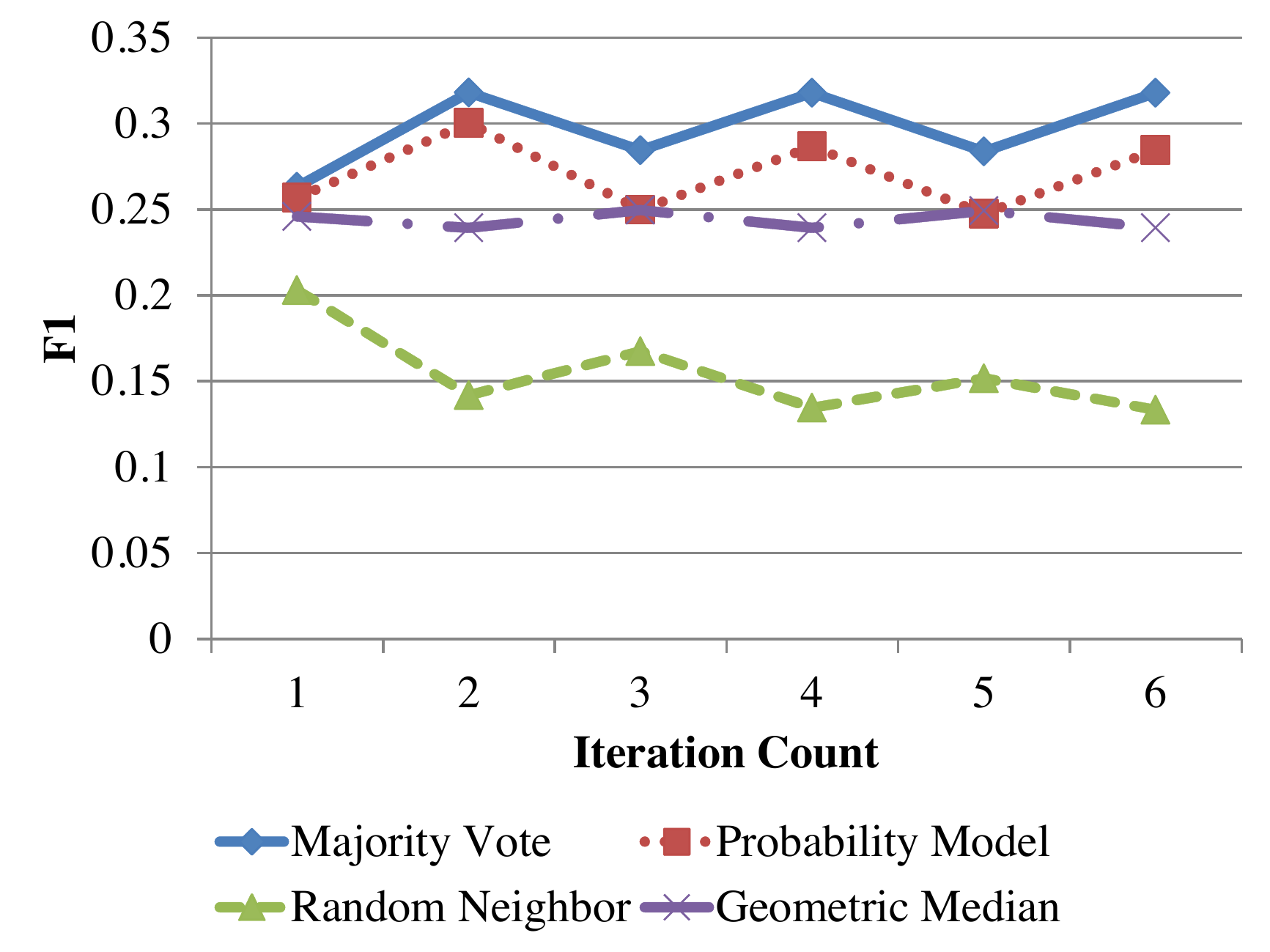}%
	\label{fig:slp-mutual-f1}%
}
\caption{The precision, recall and F1 with four select functions. The Majority Vote method has the highest precision and recall. The highest precision and recall is achieved when the iteration count is two.}
\label{fig:slp}
\end{figure}

We evaluate four select functions using the social network based on following relationships.
We show the evaluation results, which changes the select functions by changing the iteration count from one to six in Fig.~\ref{fig:slp}.
%
The Majority Vote method has the highest precision and recall in all iterations.
The Probability Model method has the second best precision and recall.
The highest precision and recall are achieved when the iteration count is two.
The Geometric Median method features small differences between the maximum and minimum values of precision and recall.
The coverage for a single iteration is 0.745,
and the coverage after two iterations is approximately 0.986.
This comparison with the select functions is not consistent with using the social network based on mention relationships reported in the previous study~\cite{Jurgens2013}.

\begin{figure}[p]
\centering
\subfloat[Mean Error Distance]{%
	\includegraphics[width=0.99\linewidth]{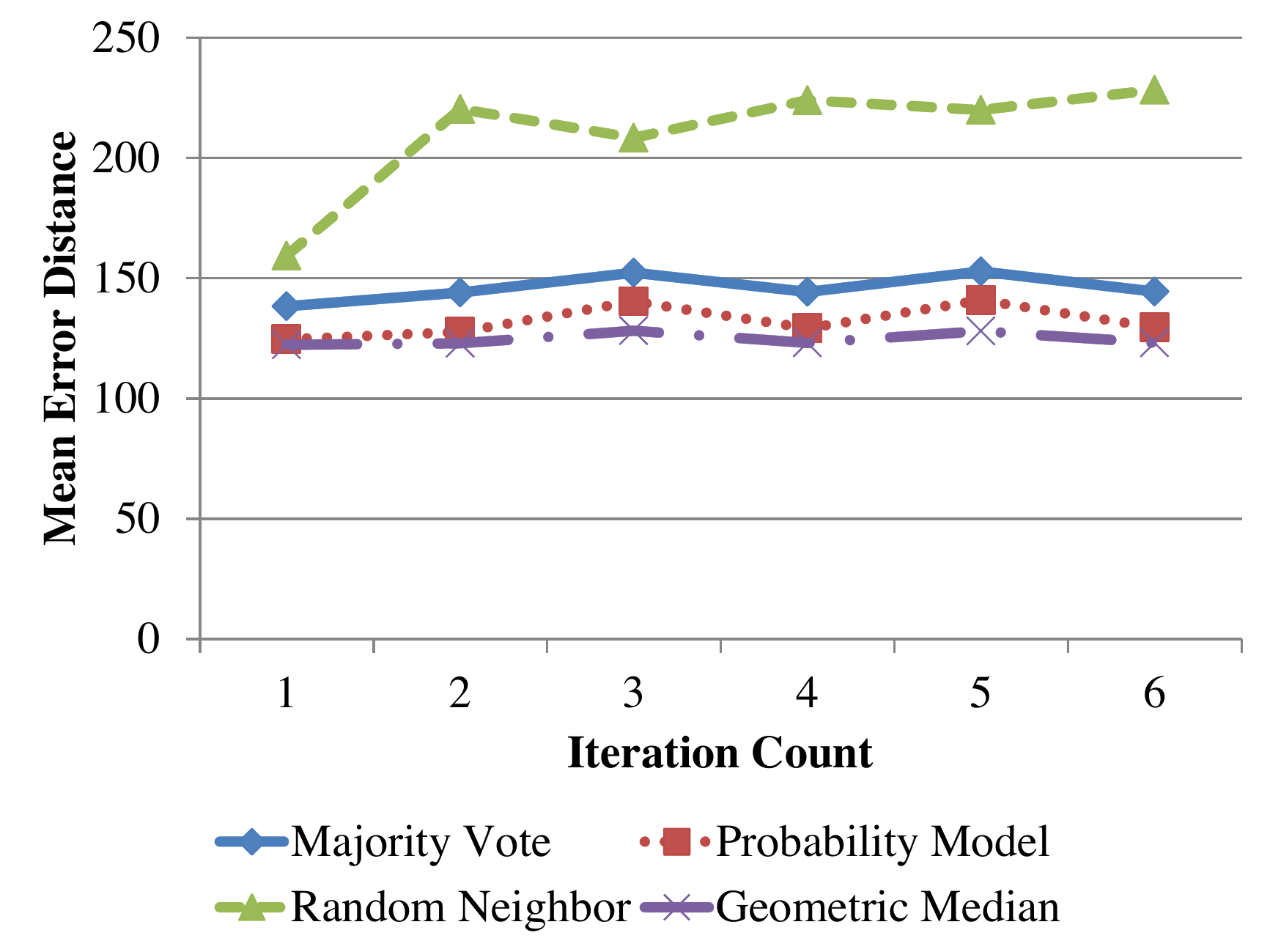}%
	\label{fig:slp-mutual-mean}%
}
\hfil
\subfloat[Median Error Distance]{%
	\includegraphics[width=0.99\linewidth]{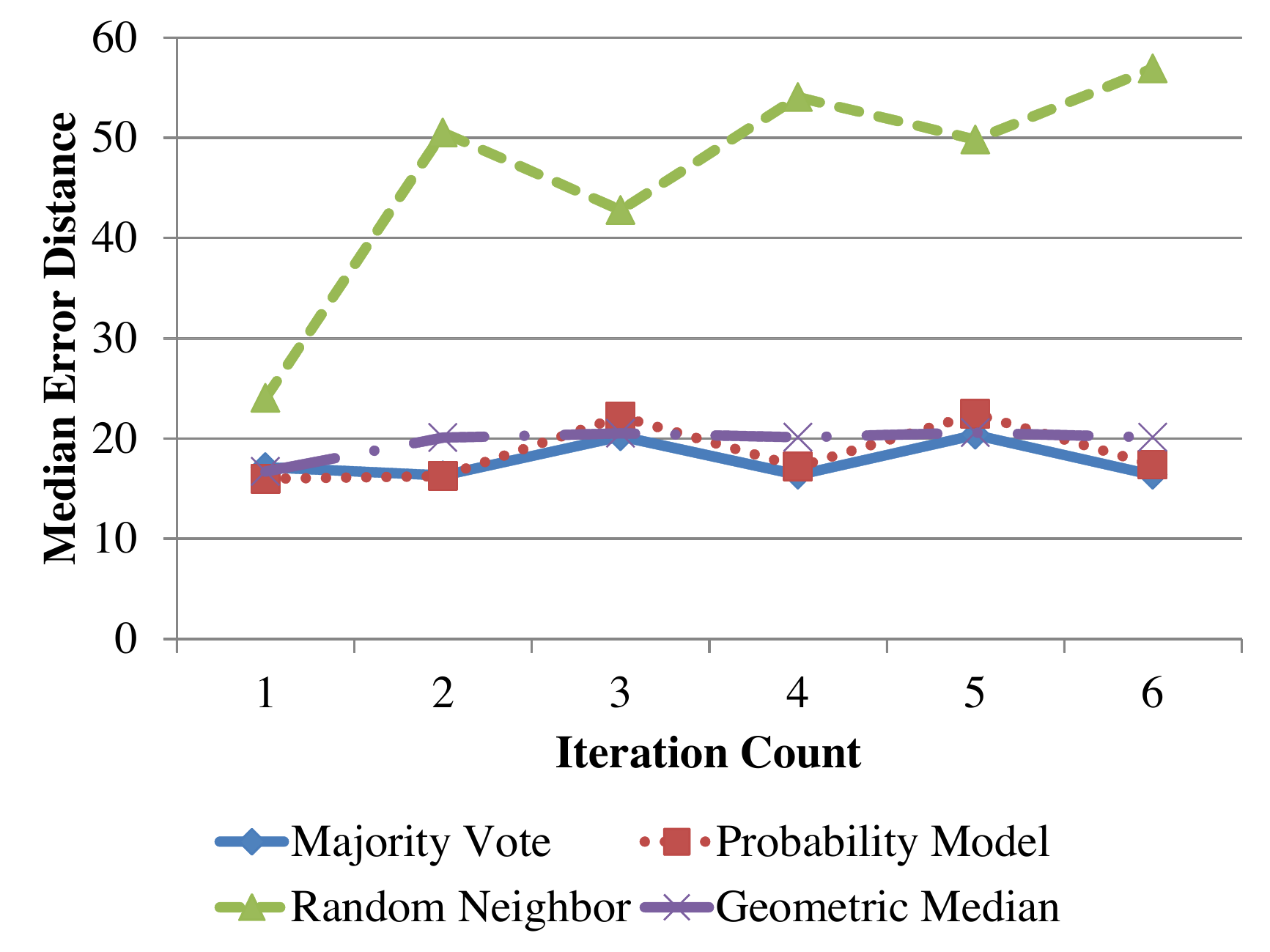}%
	\label{fig:slp-mutual-median}%
}
\caption{The mean error distance and median error distance with four select functions. The Geometric Median method has the lowest mean error distance in all iterations. The Majority Vote and Probability Model methods have lower median error distances at iteration two.}
\label{fig:slp-ed}
\end{figure}

Fig.~\ref{fig:slp-ed} shows the evaluation results by mean and median error distance.
The Geometric Median method has the lowest mean error distance in all iterations.
The Majority Vote and Probability Model methods have lower median error distances on the second iteration.

The hypothesis that the SLP improves F1 is that the estimated location can be trusted as sufficiently as the assigned home location.
The select functions can have higher precision when setting parameters, such as the number of adjacent nodes to estimate.
There is a trade-off between precision and recall.
There is a possibility that the select function having high precision can reliably estimate home location.
A detailed analysis of select functions will likely be presented in a future study.

\begin{figure}[p]
    \centering
    \includegraphics[width=0.99\linewidth]{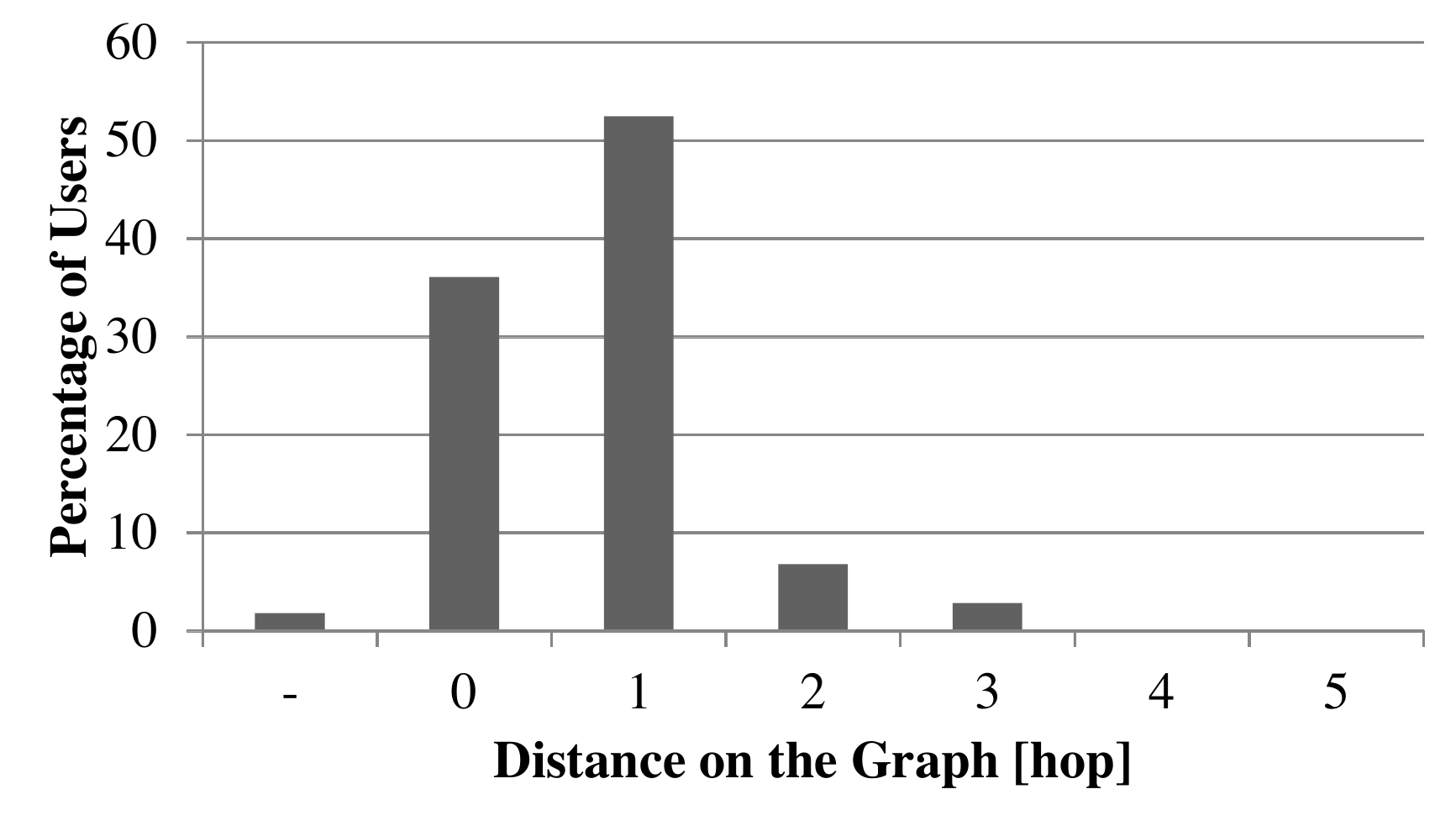}
    \caption{The distribution of the distance on the graph to the user having the same home location. ``-'' indicates the percentage of the users who are not able to reach a user that has the same home location. The distance to the adjacent node (friend) is zero-hop. Approximately 88\% of users have a user having at least one same home location within one-hop (friends and friends of friends).}
    \label{fig:cntdist}
\end{figure}

We analyze the distribution of the distance on the graph to the user having the same home location in the social network.
Fig.~\ref{fig:cntdist} shows the percentage of the users of each distance.\break
``-\nobreak'' indicates the percentage of the users who are not able to reach a user that has the same home location.
Approximately 88\% of users have a user having at least one same home location within one-hop (friends and friends of friends).
For a higher recall, to estimate the correct home location to a greater number of users is important.
The result indicates that the iteration count of SLP is sufficient with two times calculation.

\subsection{Discussion}



As indicated in Fig.~\ref{fig:slp}, the Majority Vote method had the best performance in our experiments.
Jurgens et al.~\cite{Jurgens2013} have shown that the Geometric Median method has higher performance than the other three select functions, excluding the Probability Model method, which was not used.
Two reasons can be considered for this result: the difference of the size of the home location, and the difference of the relationships in the social network.

Firstly, the home location is different in size and shape.
The Majority Vote method estimates a home location by selecting the majority area.
It is considered that the Majority Vote method is more affected by the size of the area and the shape of the area than the other methods
because the number of votes becomes the same value when the area is too small.
Therefore, a certain number of votes is necessary to conduct an accurate estimation.
We have defined a home location as an area of a city in Japan.
This definition could be a reason that the Majority Vote method had the best performance in our experiment.

Secondly, the relationship of the social network that it is based on is different.
Our study uses the social network based on following relationships, which is used in previous studies~\cite{DavisJr2011,McGee2013,Rout2013}.
By contrast, some previous studies~\cite{Jurgens2013,Jurgens2015} have used the social network based on mention relationships.
McGee et al.~\cite{McGee2011} reported that the ratio of friends located geographically near changes with the relationship between the users.
It is considered that the best select function changes through this feature of the social network.

\section{Conclusion}

We analyzed the select functions of SLP using a social network based on following relationships.
As a result, we revealed that the Majority Vote method that selects the most frequent location among the friends' locations had the highest precision and recall.
We also indicated that the iteration count of SLP is sufficient with two times calculation,
because, since 88\% of users are in the social network based on following relationships, in which users' possessed at least one correct home location within one-hop (friends and friends of friends).






\bibliographystyle{IEEEtran}
\bibliography{IEEEabrv,references}
%

%

\balance

\end{document}